# On the second harmonic generation in superlattices

G. M. Shmelev[1], E. N.Valgutskova[1], E. M. Epshtein[2]

[1]Volgograd State Pedagogical University, 4000131, Volgograd, Russia

[2]Institute of Radio Engineering and Electronics of the Russian Academy of Sciences, 141190, Fryazino, Russia

There are many works devoted to current harmonic generation in one-dimensional superlattices (SL) (see $[1-5]$). The second harmonic generation (SHG) occurs in presence of DC electric field or other cause breaking initial inversion symmetry. In present work, we investigate SHG in SL under presence of the carrier density gradient ($\nabla n$). We show that the SH intensity can be comparable with that of the current fundamental harmonic in that case.

We calculate the current along the SL axis driven by DC and AC electric fields

$$\mathbf{E}(t) = \mathbf{E}\cos\omega t + \mathbf{E}_1, \ \mathbf{E}_1 = const \tag{1}$$

and gradient $\nabla n$. All the vectors are directed along the SL axis, so that we have a one-dimensional quasi-classical problem (the fields are assumed to be non-quantizing ones). With an additive energy spectrum, the electron distribution function is $f(\mathbf{p},\mathbf{r},t) = f(\mathbf{p}_\perp, p_x, x, t)$, where $\mathbf{p}$ is electron quasi-momentum, $\mathbf{p}_\perp$ is its component perpendicular to the SL axis. For brevity, we omit $\mathbf{p}_\perp$ and write $p_x$ as $p$ below. The Boltzmann kinetic equation takes the form

$$\frac{\partial f(p,x,t)}{\partial t} + eE(t)\frac{\partial f(p,x,t)}{\partial p} + \upsilon(p)\frac{\partial f(p,x,t)}{\partial x} = St\, f(p,x,t), \tag{2}$$

where $\upsilon(p) = \partial\varepsilon(\mathbf{p})/\partial p$, $\varepsilon(\mathbf{p})$ is electron energy, $St\, f(\mathbf{p},\mathbf{r},t)$ is collision integral. In many works on the electron transport in SLs, that integral is taken within the simplest $\tau$-approximation [5]. In our case, such an approach meets with some principal difficulties. Therefore, we use a model Bhatnagar – Gross – Crooke integral

$$St f = \frac{1}{\tau}\left[ f_0 \frac{n}{n_0} - f \right], \quad (\tau = \mathrm{const}), \tag{3}$$

where $f_0 = f_0(\mathbf{p})$ is equilibrium distribution function, $\sum_{\mathbf{p}} f_0(\mathbf{p}) = n_0$ is equilibrium electron density and $\sum_{\mathbf{p}} f(\mathbf{p}, \mathbf{r}, t) = n(\mathbf{r}, t)$ is nonequilibrium one. This integral obeys necessary condition $\sum_{\mathbf{p}} St f = \sum_{\mathbf{p}} \frac{1}{\tau}\left[ f_0 \frac{n}{n_0} - f \right] = 0$. A continuity equation follows

from Eqs. (2) and (3)

$$\frac{\partial n}{\partial t} + \frac{1}{e}\mathrm{div}\, \mathbf{j}(t) = 0, \tag{4}$$

where

$$\mathbf{j}(\mathbf{r}, t) = e \sum_{\mathbf{p}} \mathbf{\upsilon}(\mathbf{p}) f(\mathbf{p}, \mathbf{r}, t) \tag{5}$$

is current density.

We seek solution of Eq. (2) with collision integral (3) as

$$f(p, x, t) = \frac{n(x, t)}{n_0} f^{(0)}(p, t) + \varphi(p, x, t). \tag{6}$$

Here the function

$$f^{(0)}(p, t) = \int_{-\infty}^{t} \exp\left( \frac{t' - t}{\tau} \right) f_0\left( p - e \int_{t'}^{t} E(t'')\,\mathrm{d}t'' \right) \frac{\mathrm{d}t'}{\tau} \tag{7}$$

obeys equation

$$\frac{\partial f^{(0)}}{\partial t} + eE(t)\frac{\partial f^{(0)}}{\partial p} = \frac{f_0 - f^{(0)}}{\tau} \tag{8}$$

and $\sum_{\mathbf{p}} f^{(0)} = n_0$ condition, while $\varphi$ is unknown function satisfying

$\sum_{\mathbf{p}} \varphi = 0$ condition.

Substituting Eq. (6) into Eq. (2) with Eqs. (4) and (8) taking into account, we

obtain an equation for $\varphi$ function:

$$\frac{\partial \varphi(p, x, t)}{\partial t} + eE(t)\frac{\partial \varphi(p, x, t)}{\partial p} + \upsilon(p)\frac{\partial \varphi(p, x, t)}{\partial x} + \frac{\varphi(p, x, t)}{\tau} = \frac{1}{n_0}f^{(0)}(p, t)\frac{\partial}{\partial x}\left[ \frac{1}{e}j(x, t) - \upsilon(p)n(x, t) \right],$$

$$(j = j_x).$$

(9)

It can be found by direct substitution that the solution of Eq. (9) is

$$\varphi(p,x,t) = \frac{1}{n_0} \frac{\partial}{\partial x} \int_{-\infty}^{t} \exp\left(\frac{t'-t}{\tau}\right) f^{(0)}\left(p - e\int_{t'}^{t} E(t'')\mathrm{d}t'', t'\right) \left[\frac{1}{e} j\left(x + \frac{\varepsilon\left(p - e\int_{t'}^{t} E(t'')\mathrm{d}t''\right) - \varepsilon(p)}{eE(t')}, t'\right) - \right.$$

$$\left. - n\left(x + \frac{\varepsilon\left(p - e\int_{t'}^{t} E(t'')\mathrm{d}t''\right) - \varepsilon(p)}{eE(t')}, t'\right) \upsilon\left(p - e\int_{t'}^{t} E(t'')\mathrm{d}t''\right)\right] \mathrm{d}t', \quad (t \gg \tau).$$

(10)

In general, the solution (10) has a formal character, because the problem in consideration is to be solved by a self-consistent way together with the Poisson equation. Besides, the inhomogeneity source should be indicated. However, such a procedure is not necessary to evaluate the SHG efficiency.

We take the electron density gradient within linear approximation below. In that case, we obtain from Eq. (10)

$$\varphi(p,x,t) = \frac{1}{n_0} \frac{\partial}{\partial x} \int_{-\infty}^{t} \exp\left(\frac{t'-t}{\tau}\right) f^{(0)}\left(p - e\int_{t'}^{t} E(t'')\mathrm{d}t'', t'\right) n(x,t') \left[\frac{1}{en_0} j^{(0)}(t') - \upsilon\left(p - e\int_{t'}^{t} E(t'')\mathrm{d}t''\right)\right] \mathrm{d}t',$$

(11)

where $j^{(0)}(t)$ is the current density calculated with substitution $f^{(0)}$ function into Eq. (5). It can be seen that Eq. (11) satisfies $\sum_{\mathbf{p}} \varphi = 0$ condition.

It has been shown in Refs. 7, 8 devoted to one-dimensional diffusion from a plane source in bulk materials in presence of a high-frequency electric field, that $n(x,t) \approx n^{(0)}(x) \ll n_0$ approximation can be assumed in real situations with $n^{(0)}(x)$ being DC component of the electron density. In that approximation, we obtain from Eq. (11) at constant temperature ($T$).

$$\varphi(p,x,t)=\frac{1}{n_0}\frac{\partial n^{(0)}(x)}{\partial x}\int\limits_{-\infty}^{t}\exp\left(\frac{t'-t}{\tau}\right)f^{(0)}\left(p-e\int\limits_{t'}^{t}E(t'')\mathrm{d}t'',t'\right)\left[\frac{1}{en_0}j^{(0)}(t')-\upsilon\left(p-e\int\limits_{t'}^{t}E(t'')\mathrm{d}t''\right)\right]\mathrm{d}t'\cdot$$

(12)

Therefore, the function (6) with $\varphi$ function defined by Eq. (12) is the desired solution of Eq. (2). Substituting the $f$ function into Eq. (5), we obtain $j(x,t)=j_0(x,t)+j_d(x,t)$, where $j_0(x,t)=j^{(0)}(t)n^{(0)}(x)/n_0$, $j_d$ being the diffusion current density

$$j_d(x,t)=-eD(t)\frac{\partial n^{(0)}(x)}{\partial x},$$     (13)

$$D(t)=\frac{1}{n_0}\sum_p\upsilon(p)\int\limits_{0}^{\infty}\int\limits_{0}^{\infty}\exp\left(-\frac{t_1+t_2}{\tau}\right)f^{(0)}\left(p-e\int\limits_{t-t_1-t_2}^{t}E(t'')\mathrm{d}t''\right)\left[\frac{1}{en_0}j^{(0)}(t-t_1)-\upsilon\left(p-e\int\limits_{t-t_1}^{t}E(t'')\mathrm{d}t''\right)\right]\frac{\mathrm{d}t_1\,\mathrm{d}t_2}{\tau}$$

(14)

is electron diffusion coefficient. Drift current density $j^{(0)}(t)$ has been calculated in Ref. 1.

The electron dispersion law in SL under tight-binding approximation takes the from

$$\varepsilon(\mathbf{p})=\frac{p_y^2+p_z^2}{2m}-\Delta\left(1-\cos\frac{p_xd}{\hbar}\right),$$     (15)

where $2\Delta$ is lowest miniband width, $d$ is SL spatial period and $m$ is in-plane electron effective mass in this case, the quasi-classical situation with the Boltzmann equation being valid is determined by following conditions: $2\Delta>>\hbar\omega$, $\hbar/\tau$, $|eE_xd|$ , $\Delta d\left|\nabla_x n^{(0)}\right|/n_0$. The diffusion coefficient takes the from

$$D(t) = \frac{2D_0}{n_0} \Big[ \sum_p f_0(p) \int\limits_0^\infty \int\limits_0^\infty \exp\left(-\frac{t_1+t_2}{\tau}\right) \sin\left(\frac{pd}{\hbar} + \frac{ed}{\hbar} \int\limits_{t-t_1-t_2}^t E(t'')\,dt''\right) \times$$

$$\times \sin\left(\frac{pd}{\hbar} + \frac{ed}{\hbar} \int\limits_{t-t_1-t_2}^t E(t'')\,dt'' - \frac{ed}{\hbar} \int\limits_{t-t_1}^t E(t'')\,dt''\right) \frac{dt_1\,dt_2}{\tau^2} -$$

$$- C_1 \sum_p f_0(p) \int\limits_0^\infty \int\limits_0^\infty \int\limits_0^\infty \exp\left(-\frac{t_1+t_2+t_3}{\tau}\right) \sin\left(\frac{pd}{\hbar} + \frac{ed}{\hbar} \int\limits_{t-t_1-t_2}^t E(t'')\,dt''\right) \sin\left(\frac{ed}{\hbar} \int\limits_{t-t_1-t_3}^{t-t_1} E(t'')\,dt''\right) \frac{dt_1\,dt_2\,dt_3}{\tau^3} \Big],$$

(16)

where $D_0 = \upsilon_0^2 \tau/2$, $\upsilon_0 = \Delta\, d/\hbar$ is the maximal electron velocity along SL axis, $C_l = \left\langle \cos\frac{l\,p\,d}{\hbar} \right\rangle$, angular brackets denote averaging over the equilibrium carrier distribution. For nondegenerate electron gas, $C_l = I_l(\Delta/kT)/I_0(\Delta/kT)$, $I_l(z)$ is modified Bessel function, $k$ is Boltzmann constant. By substituting Eq. (1) into Eq. (6) we have

$$D(t) = D_0 \Big[ \int\limits_0^\infty \int\limits_0^\infty \exp\left(-\frac{t_1+t_2}{\tau}\right) \cos\left(a\sin\omega\,t - a\sin\omega\,(t-t_1) + \frac{edE_1}{\hbar}t_1\right) \frac{d\,t_1\,d\,t_2}{\tau^2} -$$

$$- C_2 \int\limits_0^\infty \int\limits_0^\infty \exp\left(-\frac{t_1+t_2}{\tau}\right) \cos\left(a\sin\omega\,t - a\sin\omega\,(t-t_1) - 2a\sin\omega\,(t-t_1-t_2) + \frac{deE_1}{\hbar}(t_1+2t_2)\right) \frac{dt_1\,d\,t_2}{\tau^2} -$$

$$- 2C_1^2 \int\limits_0^\infty \int\limits_0^\infty \int\limits_0^\infty \exp\left(-\frac{t_1+t_2+t_3}{\tau}\right) \sin\left(a\sin\omega\,t - a\sin\omega(t-t_1-t_2) + \frac{deE_1}{\hbar}(t_1+t_2)\right) \times$$

$$\times \sin\left(a\sin\omega\,(t-t_1) - a\sin\omega(t-t_1-t_3) + \frac{deE_1}{\hbar}t_3\right) \frac{d\,t_1\,d\,t_2\,dt_3}{\tau^3} \Big]$$

(17)

where $a = \frac{deE}{\hbar\omega} \equiv \frac{E}{E_0} \cdot \frac{1}{\omega\tau}$. By expanding the integrand in Eq. (17) into Fourier series, the diffusion coefficient can be written as

$$D(t) = D^{(0)} + \sum_{s=1}^\infty D^{(s)} \cos(s\omega t - \beta_s),$$

(18)

where $D^{(0)}$ is DC component. At $\Omega = 0$ (18) does not contain even harmonics, in accordance with symmetry arguments. The analytical expressions for $D^{(0)}$ and $D^{(s)}$ coefficients, which depend on $\Omega\tau$, $\omega\tau$, $E/E_0$ and $\Delta/kT$, are rather unwieldy. So we consider some special cases.

At $\Omega\tau = 0$,

$$\frac{D^{(0)}}{D_0} = J_0^2(a) - C_2 J_0^2(a) J_0(2a) + 2\sum_{s=1}^{\infty} J_s^2(a) \frac{1}{1+(s\omega\tau)^2} -$$

$$- C_2 \sum_{\substack{s=-\infty \\ s\neq 0,\, l\neq 0}}^{\infty} \sum_{l=-\infty}^{\infty} J_s(a) J_l(a) J_{s+l}(2a) \frac{1-s(s+l)(\omega\tau)^2}{\left[1+(s\omega\tau)^2\right]\left[1+((s+l)\omega\tau)^2\right]} -$$

$$- C_1^2 \sum_{l=-\infty}^{\infty} \sum_{m=-\infty}^{\infty} \sum_{s=-\infty}^{\infty} J_l(a) J_m(a) J_s(a) \Big\{ J_{l+s-m}(a) \frac{1+s(l+m)(\omega\tau)^2 - lm(\omega\tau)^2}{\left[1+(l\,\omega\tau)^2\right]\left[1+(m\omega\tau)^2\right]\left[1+(s\omega\tau)^2\right]} -$$

$$- J_{m+s-l}(a) \frac{1-s(l+m)(\omega\tau)^2 - lm(\omega\tau)^2}{\left[1+(l\,\omega\tau)^2\right]\left[1+(m\omega\tau)^2\right]\left[1+(s\omega\tau)^2\right]} \Big\}, \tag{19}$$

where $J_s(b)$ is Bessel function of the first kind.

At $\omega\tau \gg 1$,

$$D^{(0)} = D_0 \frac{1}{1+\Omega^2\tau^2} J_0^2(a) \left[ 1 + C_2 J_0(2a) \frac{2\Omega^2\tau^2 - 1}{1+4\Omega^2\tau^2} - 4 C_1^2 J_0^2(a) \frac{\Omega^2\tau^2}{\left(1+\Omega^2\tau^2\right)^2} \right], \tag{20}$$

If AC field is absent ($a = 0$), then the result of Ref. 6 follows from Eq. (20) or (19):

$$D^{(0)} = D_0 \frac{1}{1+\Omega^2\tau^2} \left[ 1 + C_2 \frac{2\Omega^2\tau^2 - 1}{1+4\Omega^2\tau^2} - 4 C_1^2 \frac{\Omega^2\tau^2}{\left(1+\Omega^2\tau^2\right)^2} \right], \tag{21}$$

where $\Omega = E_1/E_0\,\tau$ is the Stark frequency. At $\Omega = 0$, we obtain from Eq. (20)

$$D^{(0)} = D_0 J_0^2(a) \left[ 1 - C_2 J_0(2a) \right]. \tag{22}$$

The typical behavior of the diffusion coefficient DC component describing with Eq. (19) is shown in Figs. 1($a$) and 2($a$).

It follows from Eq. (17) for second harmonic at $\Omega = 0$

$$D^{(2)} = D_0 \sqrt{g_1^2 + g_2^2}\,, \quad (\mathrm{tg}\,\beta_2 = g_2/g_1),$$

$$g_1 \equiv V_2 - C_2 V_6 - C_1^2\left(V_{10} - V_{14}\right), \quad g_2 \equiv C_2 V_8 - V_4 - C_1^2\left(V_{12} - V_{16}\right),$$

$$V_2 \equiv 2\sum_{l=1}^{\infty}[J_{l+2}(a)+J_{l-2}(a)]J_l(a)\frac{1}{1+(l\,\omega\tau)^2} \;, \quad V_4 \equiv 2\sum_{l=1}^{\infty}[J_{l+2}(a)-J_{l-2}(a)]J_l(a)\frac{l\,\omega\tau}{1+(l\,\omega\tau)^2},$$

$$V_6 = \sum_{s=-\infty}^{\infty}\sum_{l=-\infty}^{\infty}[J_{s-l-2}(a)+J_{s-l+2}(a)]J_s(a)J_l(2a)\frac{1+l(s-l)(\omega\tau)^2}{[1+((s-l)\omega\tau)^2][1+(l\,\omega\tau)^2]},$$

$$V_8 = \sum_{s=-\infty}^{\infty}\sum_{l=-\infty}^{\infty}[J_{s-l-2}(a)-J_{s-l+2}(a)]J_s(a)J_l(2a)\frac{(2l-s)\omega\tau}{[1+((s-l)\omega\tau)^2][1+(l\,\omega\tau)^2]},$$

$$V_{10} = \sum_{l=-\infty}^{\infty}\sum_{s=-\infty}^{\infty}\sum_{p=-\infty}^{\infty}[J_{l+p-s-2}(a)+J_{l+p-s+2}(a)]J_l(a)J_s(a)J_p(a)\frac{1+(3sl+sp-lp-s^2-l^2)(\omega\tau)^2}{[1+((l+p-s)\omega\tau)^2][1+(s\,\omega\tau)^2][1+(l\,\omega\tau)^2]},$$

$$V_{12} = \sum_{l=-\infty}^{\infty}\sum_{s=-\infty}^{\infty}\sum_{p=-\infty}^{\infty}[J_{l+p-s-2}(a)-J_{l+p-s+2}(a)]J_l(a)J_s(a)J_p(a)\frac{(1+sl(\omega\tau)^2)\{(l+p-s)\omega\tau\}+(l-s)\omega\tau}{[1+((l+p-s)\omega\tau)^2][1+(s\,\omega\tau)^2][1+(l\,\omega\tau)^2]},$$

$$V_{14} = \sum_{l=-\infty}^{\infty}\sum_{s=-\infty}^{\infty}\sum_{p=-\infty}^{\infty}[J_{l+s-p-2}(a)+J_{l+s-p-2}(a)]J_l(a)J_s(a)J_p(a)\frac{1-(3sl-sp-lp+s^2+l^2)(\omega\tau)^2}{[1+((l+s-p)\omega\tau)^2][1+(s\,\omega\tau)^2][1+(l\,\omega\tau)^2]},$$

$$V_{16} = \sum_{l=-\infty}^{\infty}\sum_{s=-\infty}^{\infty}\sum_{p=-\infty}^{\infty}[J_{l+s-p-2}(a)-J_{l+s-p-2}(a)]J_l(a)J_s(a)J_p(a)\frac{(1-sl(\omega\tau)^2)((l+s-p)\omega\tau)+(l+s)\omega\tau^2}{[1+((l+s-p)\omega\tau)^2][1+(s\,\omega\tau)^2][1+(l\,\omega\tau)^2]}.$$

$$(23)$$

The results describing by Eq. (23) are shown in Fig. 3 for some special cases.

The fundamental harmonic of current $j_0(x,t)$ takes the form

$$j_0^{(1)}(x,t)=[j_c n^{(0)}(x)/n_0]A\cos(\omega t-\alpha_1), \qquad (\mathrm{tg}\,\alpha_1=a_2/a_1), \qquad (24)$$

where

$$A=C_1\sqrt{a_1^2+a_2^2}\;,$$

$$a_1 \equiv 2\sum_{l=1}^{\infty}[J_{l+1}(a)+J_{l-1}(a)]J_l(a)\frac{(l\,\omega\tau)(1+(l\,\omega\tau)^2-(\Omega\tau)^2)}{[1+(l\,\omega\tau)^2+(\Omega\tau)^2]^2-4(\Omega\tau)^2(l\,\omega\tau)^2},$$

$$a_2 \equiv 2J_0(a)J_1(a)\frac{1}{1+(\Omega\tau)^2}+2\sum_{l=1}^{\infty}[J_{l+1}(a)-J_{l-1}(a)]J_l(a)\frac{1+(l\,\omega\tau)^2+(\Omega\tau)^2}{[1+(l\,\omega\tau)^2+(\Omega\tau)^2]^2-4(\Omega\tau)^2(l\,\omega\tau)^2},$$

$$(25)$$

$$j_c=e\upsilon_0\,n_0.$$

Using Eq. (25), $A$ as a function of $E/E_0$ ratio is shown in Figs. 1($b$) and 2($b$) at the parameter values corresponding to Figs. 1($a$) and 2($a$). The ratio of the current second and fundamental harmonic amplitudes is

$D^{(0)} \upsilon_0 \tau \left| \partial n^{(0)} / \partial x \right| / \left( n^{(0)} D_0 A \right)$. At $\left| \partial n^{(0)} / \partial x \right| / n^{(0)} \approx L^{-1}$ ( $L$ is diffusion length) that ratio is of order of $D^{(0)} \upsilon_0 \tau / L C_1 D_0 A$.

Let us some estimates. At $\Delta = 10^{-2}$ eV, $d = 10^{-6}$ cm we have $\upsilon_0 \approx 10^7$ cm/s. At $\tau = 10^{-12}$ s, $L = 10^{-4}$ cm, $kT \approx 2\Delta$, $\omega\tau = 0.5$, $E \approx E_0$, $\left( E_0 \approx 600 \, \text{V/cm} \right)$ the amplitudes of the fundamental and second harmonic become comparable (see Figs. 2(*b*) and 3(*b*), curves 1). At temperature $kT > 2\Delta$ situation for SHG becomes more optimal.

Similar results are obtained with the non-uniformity due to a temperature gradient.

The work was supported by the Russian Foundation of Fundamental Investigations ( Project No.02-02-16238 ).

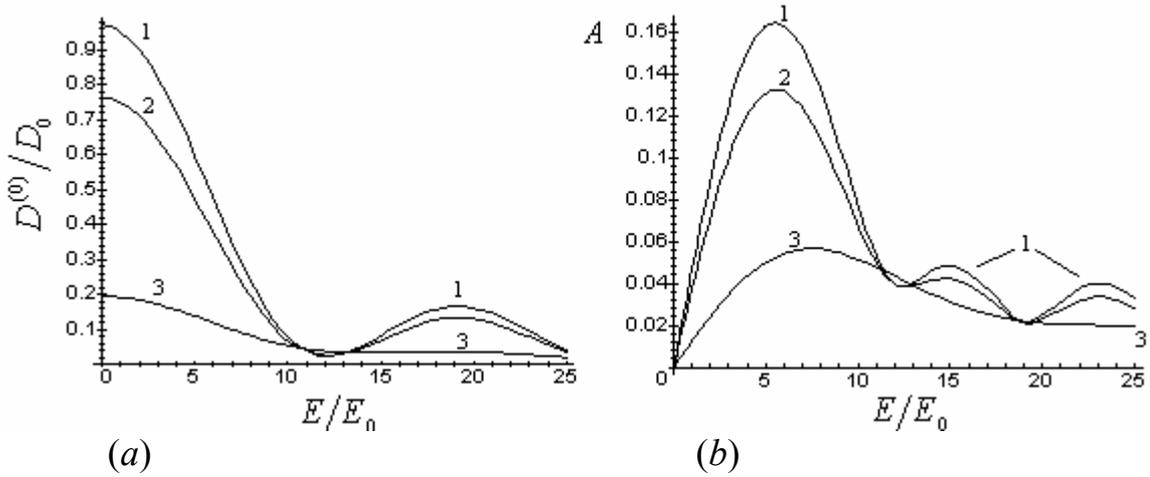

(*a*)             (*b*)

**Fig.1 *a*)** DC component of the diffusion coefficient and ***b*)** function *A* as functions of AC field amplitude $E/E_0$ at various values of $\Omega\tau$: $1 - 0$; $2 - 0.5$; $3 - 2$; ( $\Delta/kT = 0.5$, $\omega\tau = 5$ ).

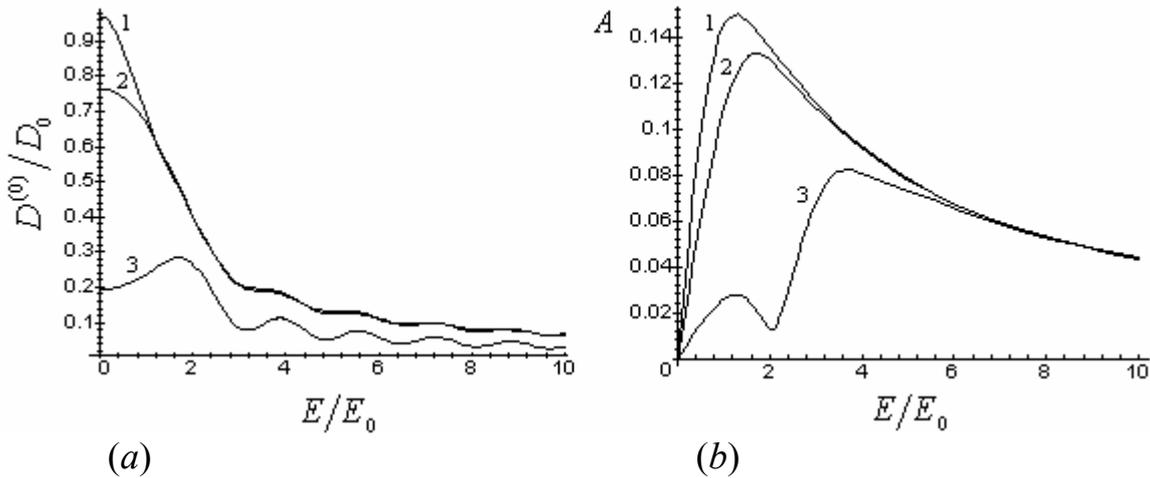

|         |         |
|:-------:|:-------:|
| *(a)*   | *(b)*   |

**Fig.2 *a*)** DC component of the diffusion coefficient and ***b)*** function $A$ as functions of AC field amplitude $E/E_0$ at various values of $\Omega\tau$: $1-0$; $2-0.5$; $3-2$; ($\Delta/kT = 0.5$, $\omega\tau = 0.5$).

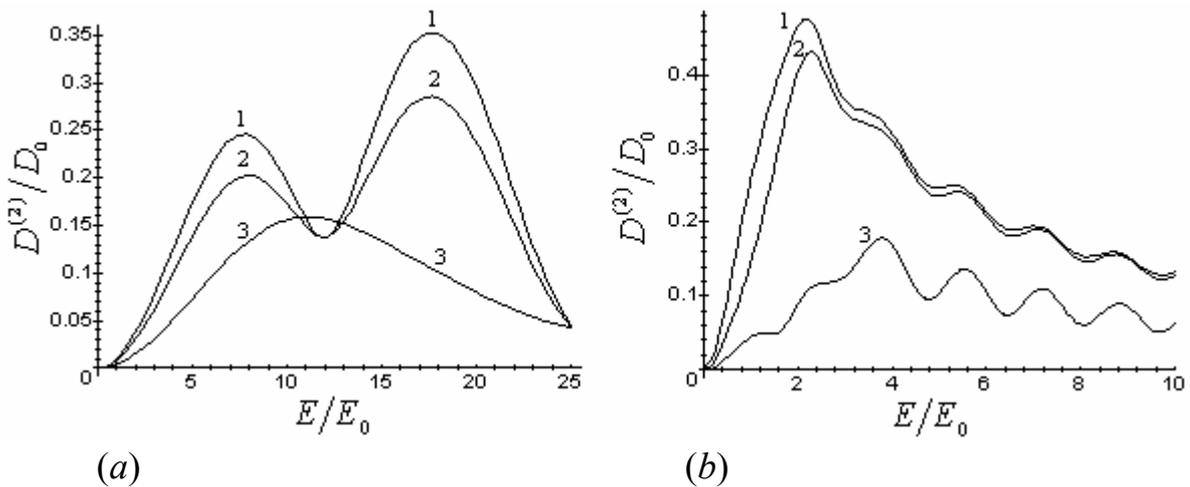

|         |         |
|:-------:|:-------:|
| *(a)*   | *(b)*   |

**Fig.3** The second harmonic amplitude of the diffusion coefficient $D^{(2)}/D_0$ as functions of AC field amplitude $E/E_0$ at various values of $\Omega\tau$: $1-0$; $2-0.5$; $3-2$; ($\Delta/kT = 0.5$, ***a***) $\omega\tau = 5$, ***b***) $\omega\tau = 0.5$).

p.156-161.